\documentclass{elsarticle}
\usepackage{amsmath,a4wide}
\usepackage{amssymb}
\usepackage{graphics}
\usepackage{graphicx}
\usepackage{subfigure}
\usepackage{xspace}
\usepackage{rotating}
\usepackage{verbatim}
\usepackage{longtable}
\usepackage{lscape}
\usepackage{multirow}
\usepackage{afterpage}
\usepackage[breaklinks]{hyperref}
\usepackage{breakurl}
\usepackage[english]{babel}
\usepackage{setspace}

\begin{document}
\setlength{\topmargin}{-12mm}

\begin{frontmatter}

\title{Identifying Clouds over the Pierre Auger Observatory using Infrared Satellite Data}

  \author{
\par\noindent
{\bf The Pierre Auger Collaboration} \\
P.~Abreu$^{61}$, 
M.~Aglietta$^{49}$, 
M.~Ahlers$^{90}$, 
E.J.~Ahn$^{78}$, 
I.F.M.~Albuquerque$^{15}$, 
I.~Allekotte$^{1}$, 
J.~Allen$^{82}$, 
P.~Allison$^{84}$, 
A.~Almela$^{11,\: 7}$, 
J.~Alvarez Castillo$^{54}$
% ...

\par\noindent
$^{1}$ Centro At\'{o}mico Bariloche and Instituto Balseiro (CNEA-UNCuyo-CONICET), San 
Carlos de Bariloche, 
Argentina \\
% ...
$^{7}$ Instituto de Tecnolog\'{\i}as en Detecci\'{o}n y Astropart\'{\i}culas (CNEA, CONICET, UNSAM), 
Buenos Aires, 
Argentina \\
% ...
$^{11}$ Universidad Tecnol\'{o}gica Nacional - Facultad Regional Buenos Aires, Buenos Aires,
Argentina \\
% ...
$^{15}$ Universidade de S\~{a}o Paulo, Instituto de F\'{\i}sica, S\~{a}o Paulo, SP, 
Brazil \\
% ...
$^{49}$ Istituto di Fisica dello Spazio Interplanetario (INAF), Universit\`{a} di Torino and 
Sezione INFN, Torino, 
Italy \\
% ...
$^{54}$ Universidad Nacional Autonoma de Mexico, Mexico, D.F., 
Mexico \\
% ...
$^{61}$ LIP and Instituto Superior T\'{e}cnico, Technical University of Lisbon, 
Portugal \\
% ...
$^{78}$ Fermilab, Batavia, IL, 
USA \\
% ...
$^{82}$ New York University, New York, NY, 
USA \\
% ...
$^{84}$ Ohio State University, Columbus, OH, 
USA \\
% ...
$^{90}$ University of Wisconsin, Madison, WI, 
USA \\
% ...
\par\noindent
(\ddag) Deceased \\
(a) Now at Konan University \\
(b) Also at the Universidad Autonoma de Chiapas on leave of absence from Cinvestav \\
(c) Now at University of Maryland \\
(d) Now at NYU Abu Dhabi \\
}

  \begin{abstract}
We describe a new method of identifying night-time clouds over the Pierre Auger Observatory using infrared data from the Imager instruments on the GOES-12 and GOES-13 satellites. We compare cloud identifications resulting from our method to those obtained by the Central Laser Facility of the Auger Observatory. Using our new method we can now develop cloud probability maps for the 3000 km$^2$ of the Pierre Auger Observatory twice per hour with a spatial resolution of $\sim$2.4 km by $\sim$5.5 km. Our method could also be applied to monitor cloud cover for other ground-based observatories and for space-based observatories.\let\thefootnote\relax\footnotetext{Corresponding author. E-mail: auger\_spokespersons@fnal.gov Phone: +49 202 439 2856.}
  \end{abstract}

\end{frontmatter}


\section{Introduction}

\begin{figure}[th]
\begin{centering}
\includegraphics[width=15cm,height=7.5cm]{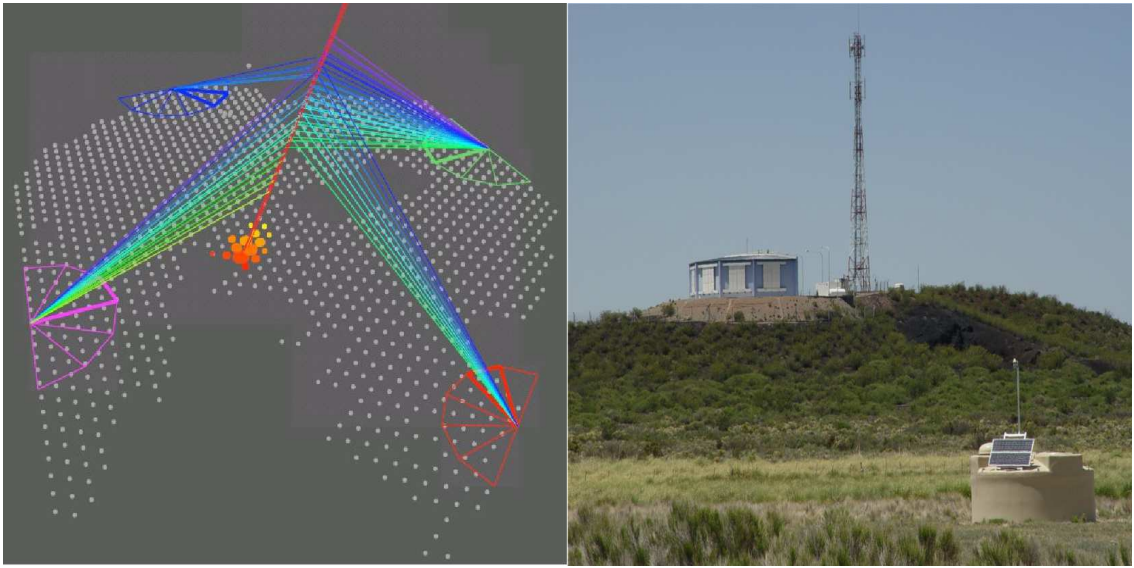}
\par\end{centering}

\caption{Left: 3D schematic of the Pierre Auger Observatory showing the four
FD stations, some SD stations and a cosmic ray event viewed by all
four FD stations. Each FD station records the development of the extensive
cosmic ray air shower comprised of billions of secondary particles.
Right: One of the 1660 SD stations in the
foreground and, atop the hill, one of the four FD stations with a
communication tower of the Pierre Auger Observatory.\label{FD}}

\end{figure}

The Pierre Auger Observatory is located in the province of Mendoza, Argentina, and covers an area of 3000 km$^{2}$. In its original layout, it detects extensive air showers
produced by cosmic rays with two different detectors (right panel of Figure \ref{FD}): a surface
detector (SD) \cite{key-3} and a fluorescence detector (FD) \cite{key-2}.
The SD consists of 1660 water-Cherenkov stations on a triangular
grid with 1.5 km spacing. Each SD station detects secondary particles
from the extensive air showers arriving at the ground. The FD system
consists of 27 air fluorescence telescopes grouped in 4 FD stations
located at the borders of the observatory. They are able to detect
fluorescence light on clear nights with low moonlight background (left panel of Figure \ref{FD}). The fluorescence light is emitted by atmospheric nitrogen through interactions with particles produced during the development of the extensive air showers in the
atmosphere. Using the fluorescence light detected by the FD,  
the longitudinal profile of the extensive air showers can be obtained. The shower
profile in turn is used to infer the energy and interaction properties
of the primary cosmic ray \cite{key-4,key-444}. 

The atmosphere influences many aspects of the generation and detection of extensive air showers. Therefore, an atmospheric group has been formed at the Pierre Auger Observatory \cite{key-13}. One of the atmospheric factors studied are the clouds. Clouds in the FD aperture
can adversely affect the measurement of shower profiles \cite{key-13}.
The Auger Observatory thus routinely employs a number of instruments \cite{key-8}:
the Central Laser Facility (CLF) \cite{key-111},
the eXtended Laser Facility (XLF), LIDARs \cite{key-7}, and IR cameras,
to identify clouds over the array. For most cosmic ray studies,
the present system is more than adequate \cite{key-5}. But information 
from a satellite can complete and enrich the ground measurements. A satellite can cover all the Pierre Auger Observatory area without interfering with the FD acquisition. While LIDARs interfere when scanning inside the FD field of view, introducing a very small dead time. A satellite cloud identification technique would supplement the ground cloud monitoring.  

Besides the standard cosmic ray showers, we are also searching for
exotic or rare phenomena. The standard cosmic ray air showers have a longitudinal development with a single well-defined shower maximum. However, a small fraction of showers has a profile that differs considerably from this average behavior and could be related to exotic or rare phenomena \cite{key-exotics}. For studying such phenomena, we need to rely heavily on well-reconstructed shower profiles (no clouds involved).  False exotic profiles may be caused by either absorption of the shower light in clouds or by side-scattering of the longitudinal Cherenkov beam within the clouds. In this way, the sensitivity to such rare events could be enhanced with a night-time cloud monitoring system that covers all the area of the array. This could be a good addition to the methods, listed in the former paragraph, that are employed to identify clouds over the array.

We have developed a method of night-time cloud
identification based on infrared-sensitive, geosynchronous satellite
observations. In this paper, we describe our method and we use atmospheric monitoring instruments at the observatory as ground truth \cite{key-131}. We were motivated to do our own cloud identification analysis using the raw satellite data, as the equivalent cloud mask product is not yet freely available.

\section{Satellite Data}

\begin{figure}[th]
\begin{centering}
\includegraphics[width=15cm,height=15cm,keepaspectratio]{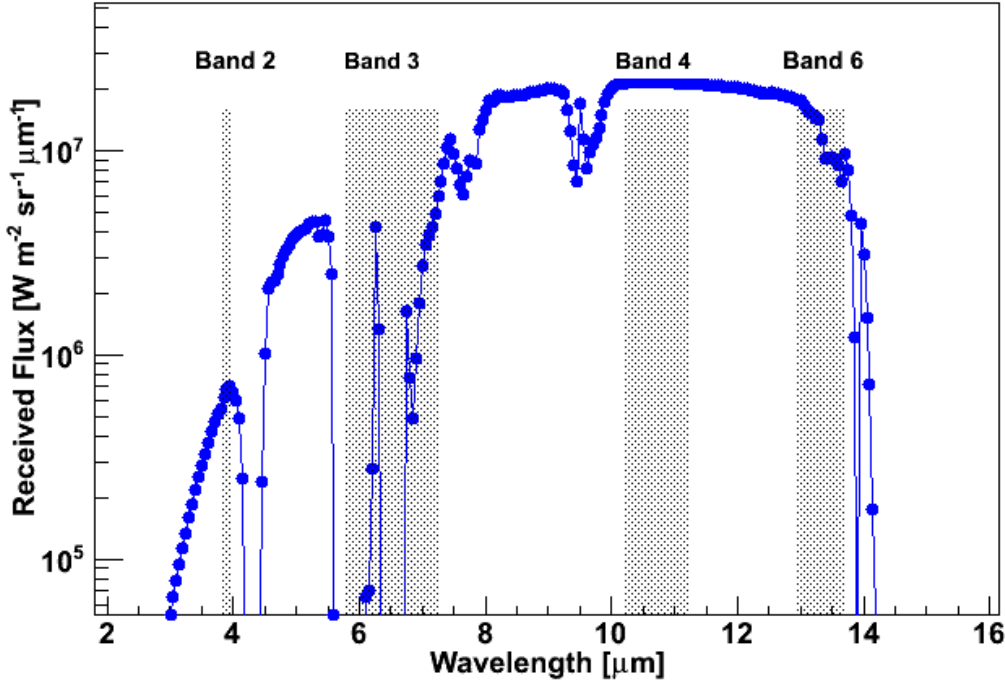}
\par\end{centering}

\caption{Wavelength band coverage for the GOES-12 Imager, superimposed on the
spectrum of a 280 K black-body. Absorption by the
atmospheric column has been applied.\label{fig:Wavelength-band-coverage}}

\end{figure}

Our analysis utilizes information provided by the Geostationary Operational
Environmental Satellites (GOES) \cite{key-1}. In particular, we use
data from the GOES-12 satellite, which was replaced by GOES-13 in April 2010. The satellite is stationed at 75 degrees West longitude. Its Imager instrument captures images of the South
American continent every 30 minutes. Full-hemisphere images are produced
in one visible band and four infrared bands, centered at wavelengths 3.9,
6.5, 10.7, and 13.3 $\mu$m. These infrared bands are labeled Band 2, 
Band 3, Band 4 and Band 6 as shown in Figure \ref{fig:Wavelength-band-coverage}. 
The bands were chosen to straddle the black-body peak for a typical range 
of Earth's surface temperatures. In Figure \ref{fig:Wavelength-band-coverage},
we show these bands superimposed on the calculated emission spectrum
for a 280 K black-body at the surface of the Earth, as viewed from
space.

The absorption effects of atmospheric water vapor are readily apparent for radiation in Band 3, and to
a lesser extent in Band 6. Radiation in Bands 2 and 4 is essentially
unaffected by passage through the atmospheric column. 

Each pixel in the infrared band has a nadir resolution of 4 km x 4 km. When projected on the ground at the Pierre Auger Observatory, the distance between the center of each pixel is about 2.4 km longitudinally and 5.5 km latitudinally. The pixels become oversampled longitudinally. The location of each pixel within the data stream is completely identified. All information can be separated and reformatted on the ground. Thus, the overlapping regions are removed and only the relevant information for the 2.4 km wide pixel is kept. The visible band resolution is higher. 

Each
pixel in each wavelength band contains the latitude
and longitude of the pixel center and the uncalibrated radiance. Each
uncalibrated radiance is subsequently transformed to a calibrated
radiance for a particular channel and detector as described by the National Oceanic and Atmospheric Administration - (NOAA) \cite{conversion}. 

\begin{figure}[th]
\begin{centering}
\includegraphics[width=10.6cm,height=10.6cm]{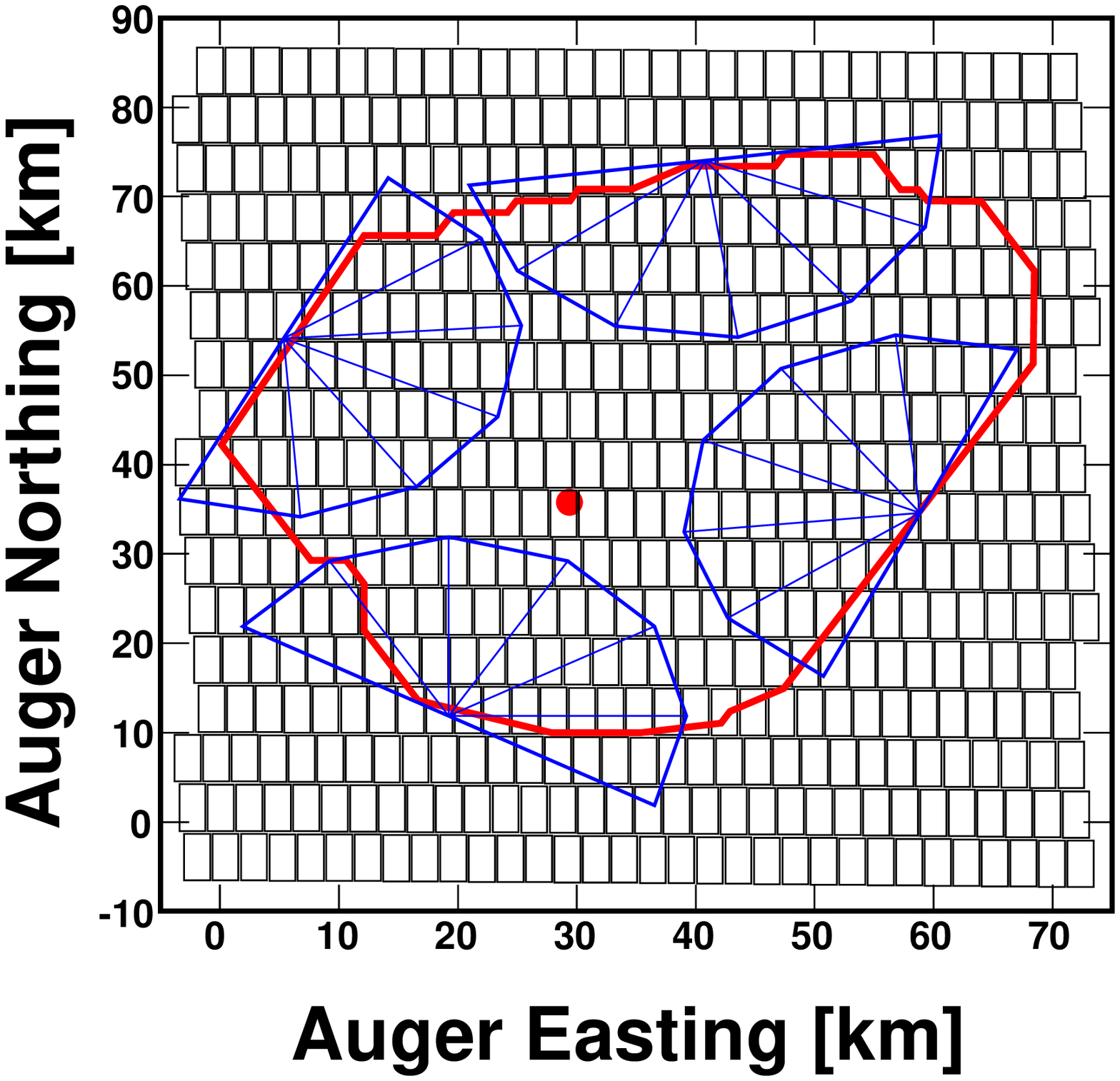}
\par\end{centering}

\caption{Pixelation of the satellite data. Shown are the borders of the SD (red thick lines), the field of view of the FD telescopes (blue thin lines) and
the CLF (red circle). The origin of this map corresponds to 6060000
 Northing and 440000 Easting from zone 19 in UTM coordinates.\label{fig:Pixelation-of-the}}

\end{figure}

The archived raw data are publicly available from the NOAA website \cite{noaawebsite}.
For our study, we selected a rectangular region centered at the Pierre
Auger Observatory (S 35.6$^\circ$, W 69.6$^\circ$). We restricted our analysis to the
infrared band data, as we are only interested in night-time cloud cover
information. Data for each observation period arrives in 4 binary-formatted files, one for each infrared wavelength band. These files
contain information for 539 pixels shown in Figure \ref{fig:Pixelation-of-the}. 

We have written analysis routines which read these four binary-formatted files
and convert the radiance for each pixel to brightness temperatures
T2, T3, T4 and T6. Here the integer labels the corresponding wavelength
band. If $I_{B_{\lambda}}$ is the measured radiance at a given wavelength
$\lambda$, then the brightness
temperature is given by:

\begin{equation}
 T_{B_{\lambda}} = \left(\frac{hc}{k\lambda}\right) \left( \ln\left(\frac{2hc^{2}}{I_{B_{\lambda}}\lambda^{5}}+1\right)\right)^{-1}.
\end{equation}
 
Recall that $T_{B_{\lambda}}$ should equal the actual temperature
$T$ if the emitting surface was a perfect black-body. For real emitting
surfaces, the brightness temperature is smaller than the actual temperature, as the measured
radiance $I_{B_{\lambda}}=\epsilon B_{\lambda}$ , where $B_{\lambda}$
is the black-body radiance given by the Planck function and $\epsilon$,
the emissivity, is less than unity for a real emitting surface. Brightness temperatures thus
vary with both the temperature and emissivity of the emitting surface.
The brightness temperatures associated with a given pixel are
the basic quantities from which cloud determinations are made. The uncertainty for GOES-12 Imager bands is less than 0.2 K at 300 K.

\section{Cloud Identification Principles}

Clouds are generally colder than the surface of the Earth. Brightness
temperatures obtained in the non-absorbing infrared bands, T2 and
T4, should consequently be lower for cloud-covered pixels. Precipitous
drops in the value of either T2 or T4 should act as an indicator for
the presence of a cloud. 

Clouds are not pure black-body emitters at infrared
wavelengths. Typically, they have low emissivities compared to
the nearly black-body emitting Earth. This has the effect of further
lowering the measured brightness temperatures T2 and T4 for cloudy
pixels.

There is a wavelength dependence in the emissivity
of cloud surfaces, which is much greater than that for the surface of the Earth. This dependence arises
because the depth into which one can receive radiation from a cloud
depends on the relationship between cloud droplet size and wavelength.
The quantity T2-T4 is sensitive to emissivity differences between
the two bands, but not to the overall temperature, as both T2 and T4 respond
to the temperature equivalently. One thus expects T2-T4 to be larger
for clouds than for the surface of the Earth.

Considering that clouds consist of a mixture of water vapor and
liquid water droplets, clouds can also modulate the absorption of radiation at 6.5 $\mu$m (Band 3). As we can see in Figure \ref{fig:Wavelength-band-coverage}, this band is the most sensitive to water vapor. In this way, the brightness temperature T3 can vary with
the fraction of cloud in a pixel. 

Cloud identification algorithms employing combinations of the brightness
temperatures T2, T4 and T3 appear promising.

\section{Cloudy/Clear Pixel Tagging with the CLF and FD}

\begin{figure*}[th]
\begin{centering}
\includegraphics[width=10.5cm,height=10.5cm,keepaspectratio]{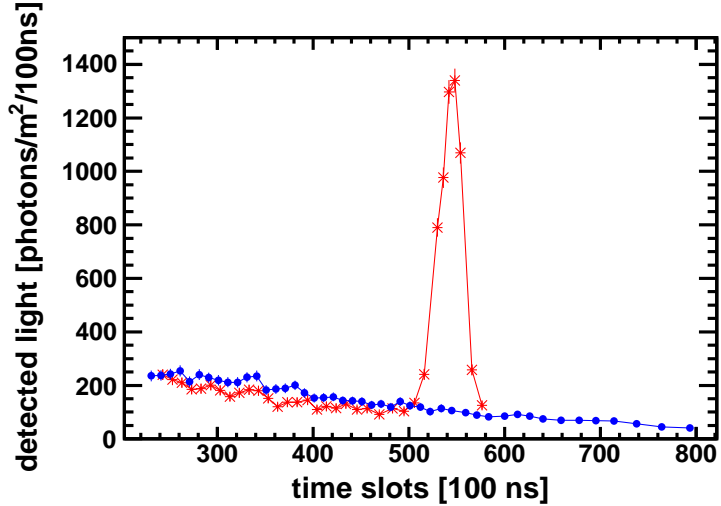}
\par\end{centering}

\caption{CLF vertical laser profile as seen from the FD station
at Los Leones during a clear night (smooth profile with blue circles). A reflection from a cloud layer immediately above the CLF shows up as a peak on the CLF vertical laser profile (red star profile with its peak).\label{fig:CLF-vertical-laser}}

\end{figure*}

It is possible to test the efficacy of the algorithms for cloud identification by checking the cloudy/clear state of the
pixel encompassing the CLF (CLF pixel). Every
15 minutes, while the FD operates, the CLF
produces a series of 50 vertical laser shots which are observed by
all four FD stations. The FD detects the presence of clouds in the
vicinity of the CLF as distortions in the otherwise smoothly falling light profiles (blue circles in Figure \ref{fig:CLF-vertical-laser}). Clouds immediately above and in the path of the CLF laser beam show up as peaks in the light profiles due to direct scattering, whereas clouds between
the FD stations and the CLF show up as drops due to absorption. The latter may not actually be located within the CLF pixel. We ignore clouds identified by dips and include only clouds identified by clearly observed peaks in this study, as they can more unambiguously
be compared with satellite measurements of the the CLF pixel. A typical CLF vertical laser shot profile indicating the presence
of a cloud layer above the CLF is shown in Figure \ref{fig:CLF-vertical-laser} (profile with peak with red stars).
The field of view of the FD restricts the maximum height of detected
cloud echoes to less than 14 km. 

We associate a smooth CLF profile with a ``clear CLF'' state 
and a profile containing a peak with a ``cloudy CLF'' state. We carefully observe each CLF profile and select only those that are reasonably smooth or contain  an obvious peak. We discard the profiles that are difficult to define. Typically, each satellite image is bracketed in time by two CLF shots,
one 9 minutes before and the other 6 minutes after the timestamp of the satellite image. The CLF pixel
is tagged as ``clear CLF pixel'' (``cloudy CLF pixel'') if the
two bracketing CLF profiles were both identified as ``clear CLF'' (``cloudy CLF'') states. This is to mitigate the effects of short-term cloud cover changes. The data used in this study were obtained over the period of a year in 2007. 

As was mentioned in Section 1, the Pierre Auger Observatory employs three cloud identification instruments: the CLF, the cloud cameras, and the LIDAR system.
We chose the CLF to do our ground truth study because its observations best match the geometry and time frame of the satellite observations. Results from the cloud cameras and satellite are not unambiguously comparable as the two devices detect clouds from very different geometric perspectives. The LIDAR system and satellite share some geometric perspective but because of the manner in which the LIDAR data is presently obtained and analyzed observations are not easily matched in time and space.  

\section{Ground Temperature Correlation}

As explained in Section 3, the brightness temperatures T2 and T4 are each
equally sensitive to the temperature of most of the emitting surfaces framed
by a pixel. For clear pixels, T2 and T4 should be correlated with
the temperature of the Earth's surface. Under clear conditions,
the brightness temperature of the ground in these bands is very nearly
equal to the actual temperature, as the emissivity of the ground is
slightly smaller than unity. 

\begin{figure}[th]
\begin{centering}
\includegraphics[width=14cm,height=14cm,keepaspectratio]{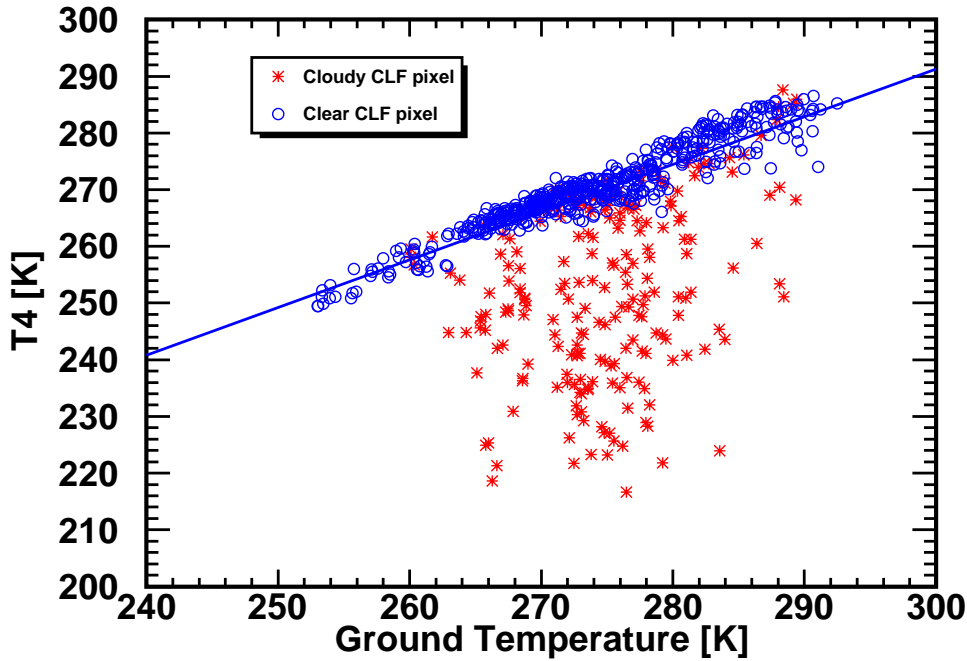}
\par\end{centering}

\caption{Brightness temperature T4 vs. ground temperature of the CLF pixel in
2007. Tagged ``cloudy CLF pixels''
are plotted as red stars and tagged ``clear CLF pixels''
as open blue circles. The blue line is the fitted line for the tagged ``clear CLF pixels''. \label{fig:Brigtness-temperature-T4}}

\end{figure}

We are able to test this relationship for the CLF pixel, where
the ground temperature has been regularly recorded by a weather station installed 2 m above the ground.
In Figure \ref{fig:Brigtness-temperature-T4}, we plot values of the
brightness temperature T4 of the CLF pixel vs. the ground temperature
at the CLF for data taken while the FD was operating in 2007.  In this figure, tagged ``cloudy CLF pixels''
are plotted as red stars and tagged ``clear CLF pixels''
are designated as open blue circles. There is an evident correlation between brightness temperature and
ground temperature for the tagged ``clear CLF pixels''. The corresponding fitted line for the tagged ``clear CLF pixels'' is also shown in Figure \ref{fig:Brigtness-temperature-T4}. The intercept of the line is 38.9 K and the slope is 0.84. This is expected, as the Earth's surface does not have a perfect unity emissivity.  

This combination of satellite and ground observables would make a
nice cloud identifier, if the ground temperature were precisely known at
each of the ground pixels throughout the array. Unfortunately, ground temperatures
are recorded only at 5 locations within the 3000 km$^{2}$ area. In Figure \ref{fig:Brigtness-temperature-T4}, we can observe that the clear region is very narrow and that there is also a small overlap between the clear and cloudy regions. If we were to use interpolations of the temperature from the 5 locations of the array to infer the temperature for each pixel, we would get larger uncertainties. This might not be a problem for the broader cloudy region, but it would be critical for the clear region and the overlap region. Thus, we proceeded to develop
a cloud identification method based on satellite-derived brightness
temperatures alone.

\section{Satellite-Based Cloud Identification}

\subsection{Ground-truthing with CLF/FD system}

\begin{figure}[!h]
\includegraphics[width=15.6cm,height=7.5cm]{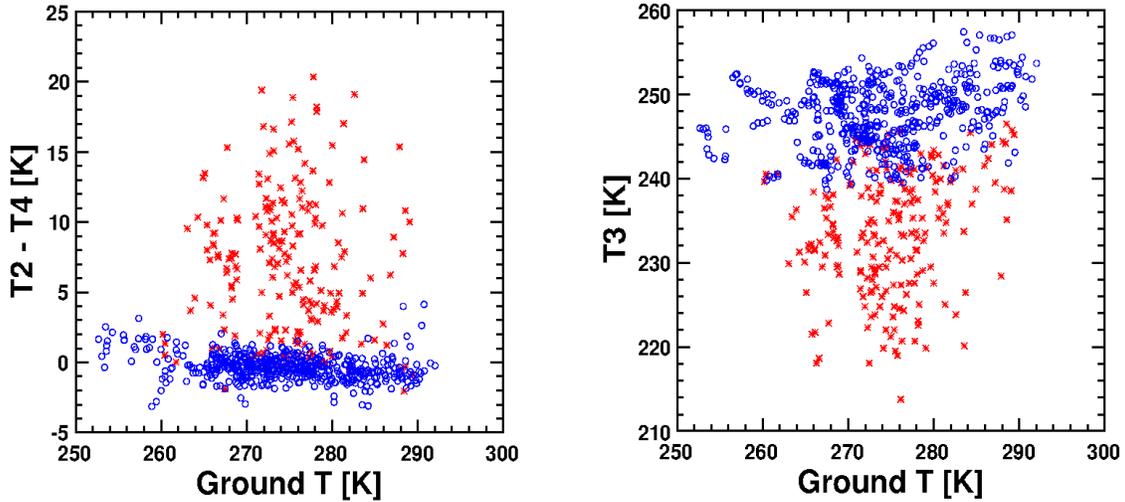}
\caption{Left: brightness temperature difference (T2-T4) vs. ground temperature
of the CLF pixel in 2007. Right: Band 3 brightness temperature (T3)
vs. ground temperature of the CLF pixel in 2007. Open blue circles (red stars) were tagged ``clear CLF pixels'' (``cloudy
CLF pixels'') as determined from the CLF study.\label{fig:t2minust4groud}}

\end{figure}

\begin{figure}[!h]
\begin{centering}
\includegraphics[width=12cm,height=12cm,keepaspectratio]{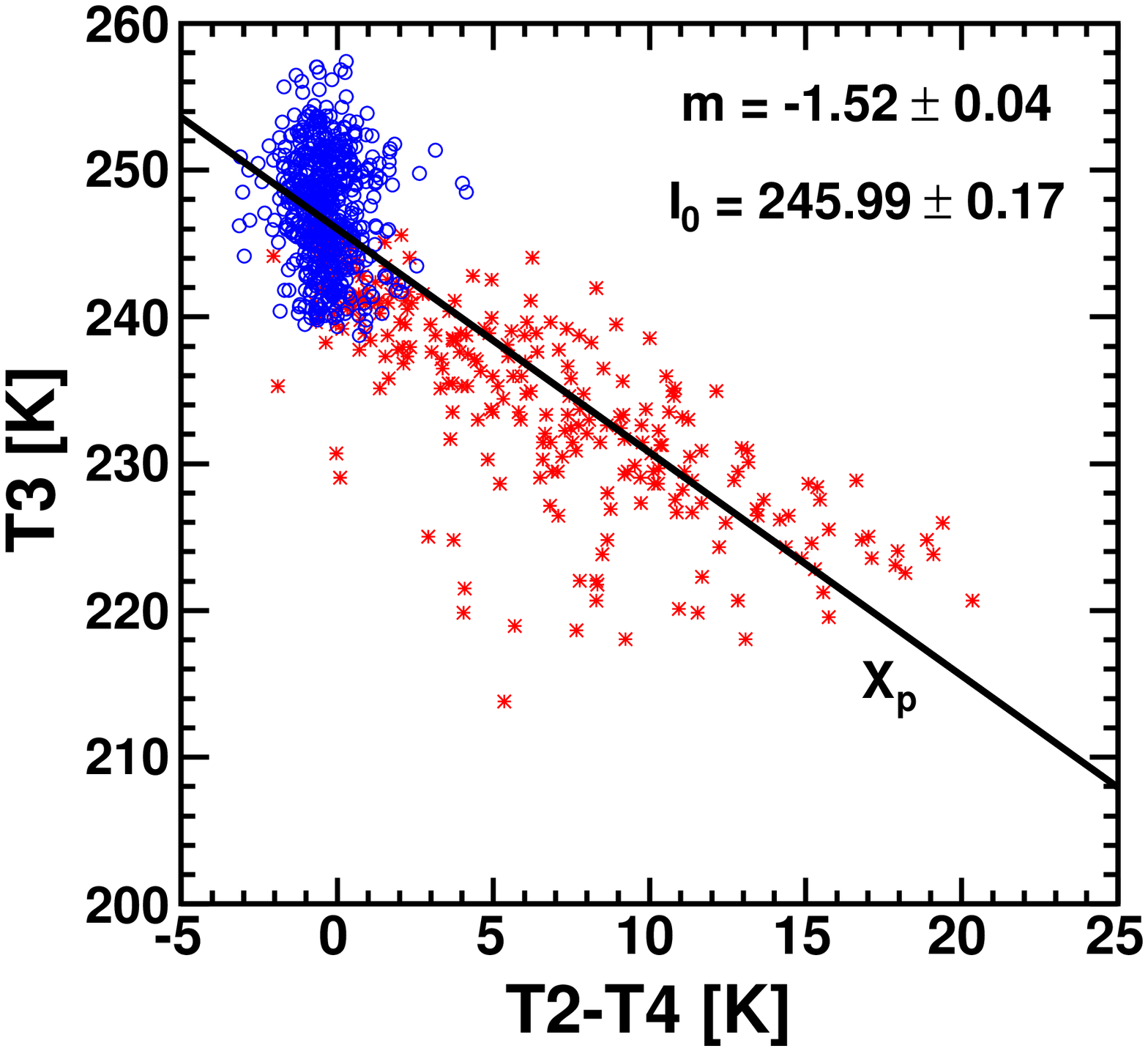}
\par\end{centering}

\caption{Brightness temperature T3 vs. brightness temperature difference T2-T4 of the CLF pixel in 2007. Open blue circles (red stars) were tagged ``clear CLF pixels'' (``cloudy
CLF pixels'') as determined from the CLF study. $I_{0}$ is the value when T2-T4$ = 0$  and $m$
is the slope of the fitted line. $X_p$ is the principal axis of the fitted line.  \label{fig:Brightness-temperature-difference}}

\end{figure}

We have identified two satellite-based quantities that appear to distinguish between
``clear CLF pixels'' and ``cloudy CLF pixels''. These are the difference
between the two unattenuated brightness temperatures (T2-T4) and the
highly attenuated brightness temperature (T3). Both are only mildly
dependent on the ground temperature (see Figure \ref{fig:t2minust4groud}), minimizing the dependence of our method
on seasonal, weekly or daily temperature variations.  
 The use of either satellite-based quantity by itself would appear to do as well at cloud discrimination
as the T4 vs. Ground Temperature method described in Section 5. However,
a combination of these quantities should perform even better. In Figure
\ref{fig:Brightness-temperature-difference}, we
plot T3 vs. T2-T4 for the CLF pixel data from 2007. The tagged ``clear CLF pixels'' (open blue circles) congregate
in the upper left quadrant of the plot. The tagged
``cloudy CLF pixels'' (red stars) form an anti-correlated linear feature. The two populations are well-separated when plotted in these variables. 


%

In an effort to maximize the discriminating power of the method we
project the data from Figure \ref{fig:Brightness-temperature-difference}
on to the principal axis $X_p$ defined by a line fitted to the overall
distribution. We use the rotation $X_{p}=(T_{2}-T_{4})\cos\theta-(T_{3}-I_{0})\sin\theta$. Here $I_{0}$ is the value when T2-T4$ = 0$ for the fitted line in
Figure \ref{fig:Brightness-temperature-difference}, 
and $\theta=arctan(m)$, where $m$ is the slope. 

\begin{figure}[!h]
\includegraphics[width=15.6cm,height=7.5cm]{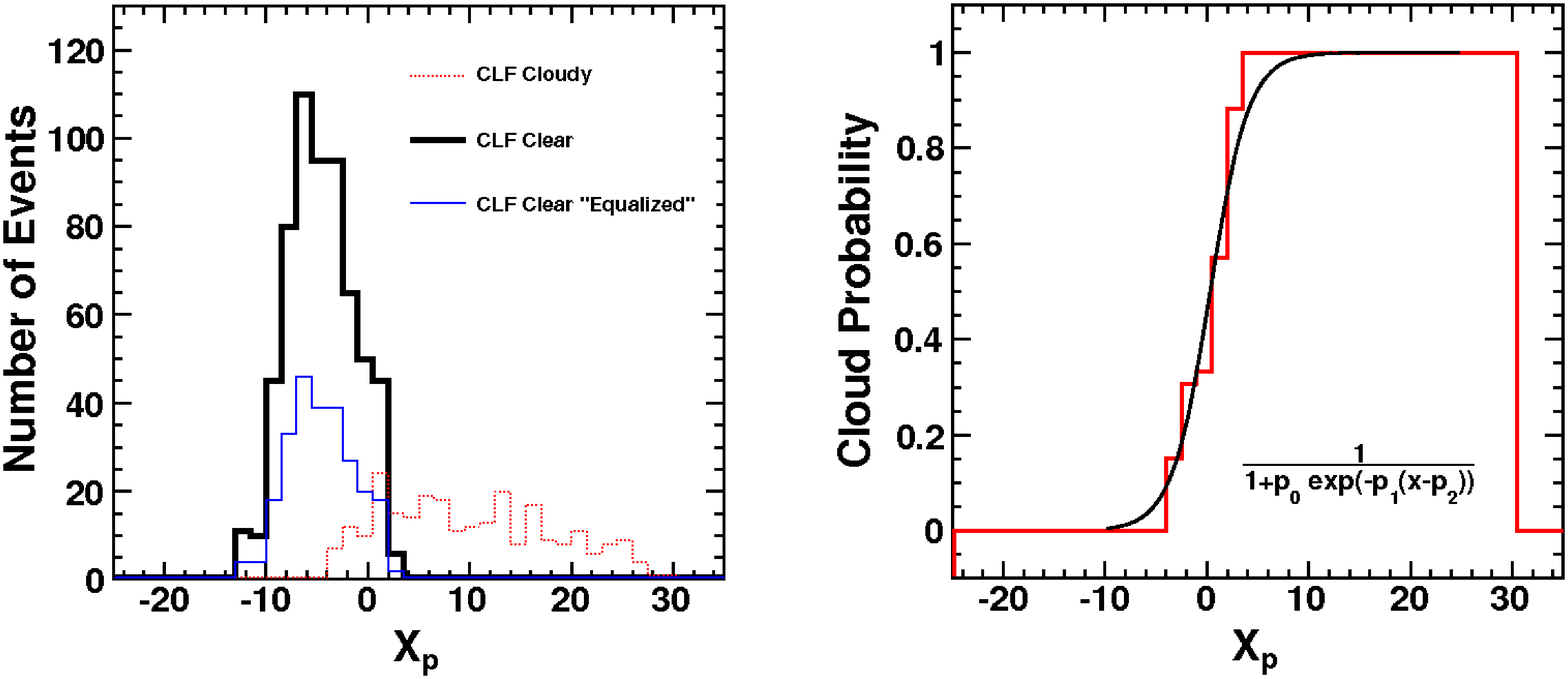}

\caption{Left: clear (black thick line), clear ``equalized'' (blue thin line on the left), and cloudy
(red dashed line on the right) tagged distributions in rotated (principal axis  $X_p$ ) system. Right: cloud
probability histogram with fitted empirical function. \label{fig:empirical function} }

\end{figure}

On the left of Figure \ref{fig:empirical function}, we show one-dimensional histograms of $X_{p}$ for the clear (black thick line) and cloudy (red dashed line on the right) tagged data. Also shown is a clear
pixel ``equalized'' histogram (blue thin line on the left) scaled to have the same area as the cloudy pixel histogram.
Suitably normalized, these histograms represent probability density
functions for $X_{p}$ conditional on the cloudy or clear state of the pixel. Using information
from the equalized clear histogram and the cloudy histogram, we
calculate a cloud probability for each bin in $X_p$ by
dividing the number of cloudy entries by the sum of the cloudy and
clear entries. The resulting distribution of cloud probability versus $X_{p}$ is shown on the right of Figure \ref{fig:empirical function}.
Also shown is a functional representation of the distribution. 

The separation along the principal axis is not perfect. There is a small overlap in the $X_{p}$ distributions shown   
on the left panel of Figure \ref{fig:empirical function}. We will discuss the possible reasons for the overlap in Subsection 6.3.

\subsection{Maps of Cloud Probability and their Application to the Auger Observatory}

\begin{figure}[!h]
\begin{centering}
\includegraphics[width=15cm,height=18cm,keepaspectratio]{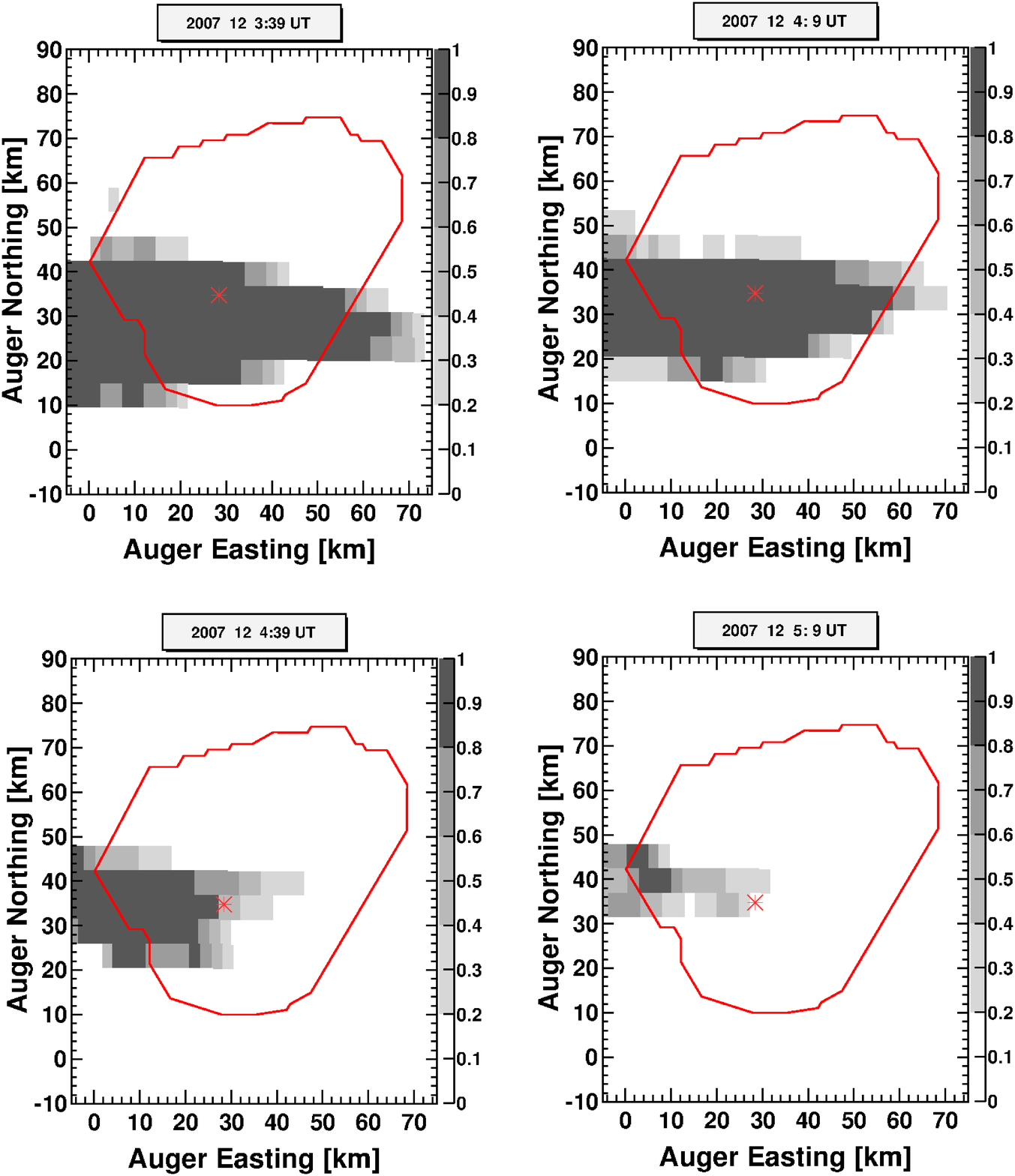}
\par\end{centering}

\caption{Examples of cloud probability maps. Shown are four successive cloud
probability maps of the Pierre Auger Observatory. The
progress of a cloud layer can be seen as it moves from East to West. Pixels  are colored in accordance with the gray scale to the right of
the maps. Shown are the borders of the SD (red) and the location of the CLF (red star).\label{fig:cloudmaps-1}}

\end{figure}

Based on the cloud identification scheme just described, we have generated a collection of cloud probability maps covering the Pierre Auger Observatory during the course of its operation. In doing this we have assumed that there is nothing special about the CLF pixel in regards to the satellite-based cloud identification algorithm. The cloud probability for each pixel was determined by evaluating the empirical function given in Figure \ref{fig:empirical function} at its particular value of $X_{p}$. 
 As an example, a cloud
bank moving to the West is shown in the sequence of four cloud probability maps at 30 minute intervals in Figure \ref{fig:cloudmaps-1}. Cloud probabilities for each pixel are plotted according to the
gray scale defined at the right of the cloud probability maps. 

%

Maps were generated for each satellite image available from all the FD running nights since 2004. In general, one image was available every 30 minutes. In addition, nightly animated maps were constructed. These maps (especially 
the animated versions) are useful in visualizing the cloud cover during and around the occurrence of interesting cosmic ray events. 
In particular, this helps in distinguishing between cloud-distorted shower profiles and those corresponding to exotic events.  

The array of cloud probabilities and timing information additionally are stored in the Auger GOES database for further reconstruction analysis of the cosmic ray data. 

We have developed a simple routine to be used within the Auger offline \cite{off} analysis framework to extract cloud-cover information from the GOES database. At present the code merely extracts the cloud probability for a pixel given a coordinate and a timestamp. In the future we anticipate the development of a more sophisticated routine that will provide the final user with the cloud state of all the pixels between the FD and the shower path. With this information we hope to recover some fraction of the data conservatively thrown away on marginally cloudy nights. We also hope to use it for a fast veto of false exotic events.

The Advanced Data Summary Trees (ADST) files, which are used in high-level Auger data analysis, contain cloud information which comes from the LIDAR systems and recently information from the GOES database and IR cloud cameras has been added.      

\subsection{Reliability of the method }

 In the previous section it was pointed out that there exists a small overlap between the $X_{p}$ distributions for ``clear CLF pixels''  and ``cloudy CLF pixels''. Pixels with $X_{p}$ falling within the overlap region thus have an ambiguous cloud status. In this section we discuss some possible reasons for this overlap. Since we used a full year of data, this overlap should include almost all the possible cases.

The CLF data and the Satellite images are not obtained simultaneously. They may be offset in time by 6 or 9 minutes. If the cloud cover in the CLF pixel changes significantly on this time scale, the two instruments will inevitably disagree on the cloud status of the pixel. The result is a spreading of both clear and cloudy distributions into the intervening region. However, the probability of this is low as we always used satellite images which were time-bracketed by two unchanging CLF measurements. 

The overlap region may also result from a spreading of the satellite ``cloudy'' data into the CLF ``clear'' distribution. This occurs when the satellite is more cloud-sensitive than the CLF system. This can happen due to the mismatch in geometric perspective between the satellite and CLF system. The CLF laser shots probe the cloudy/clear state of only a small portion of the pixel as seen from the satellite. The laser beam illuminates an area less than 100 m across whereas the pixel itself may measure several kilometers on a side. This effect may be important when the cloud cover is non-uniform but will disappear under overcast conditions. Another contribution may arise from the fact that the field of view of the FD is such that clouds whose bottom surfaces are above 14 km are undetectable by the CLF system.  However, clouds at such heights are rare.

We also expect a distribution overlap in cases where the CLF is more cloud-sensitive than the satellite. This is the situation for optically thin clouds. Thin clouds produce a negligible change in the overall infrared flux emitted by a pixel rendering them undetectable by the satellite. Optically thin clouds are not important for distortions due to absorption, but could indeed act as side-scatterers. By averaging the profiles of up to 50 laser shots, the CLF system can readily detect thin-cloud echoes above the night-sky background.  The presence of thin clouds have the effect of spreading the satellite ``clear'' data into the CLF ``cloudy'' distribution as the CLF detects clouds invisible to the satellite. 
 
In principle, by the above means the CLF/satellite system can discriminate between thin and thick clouds. Thin clouds should frequently be seen by the CLF alone while thick clouds are seen by both monitoring instruments. However the thin clouds detected in this way would have not always a discernible effect on the profiles of single naturally occurring cosmic ray showers as they could only have been detected through the average of many laser-simulated single showers. We would like to remark that our goal is cloud identification and not identification of the cloud type.

As a sanity check for our cloud probabilities, we can inspect our cloud probability maps. For example in Figure \ref{fig:cloudmaps-1}, we can see that the pixels with high cloud probabilities are continuous and also are commonly surrounded by pixels with lower cloud probabilities. 

There is some uncertainty in the ground coordinates of the pixel centers. This uncertainty, if sufficiently large could lead to a mis-identification of the CLF pixel and result in the satellite and CLF monitoring the cloud content of two different pixels. The spatial uncertainty in the satellite pixel location could contribute also to the overlap shown on the left panel of Figure \ref{fig:empirical function}.
     
\begin{figure}[!h]
\begin{centering}
\includegraphics[width=10.5cm,height=10.5cm,keepaspectratio]{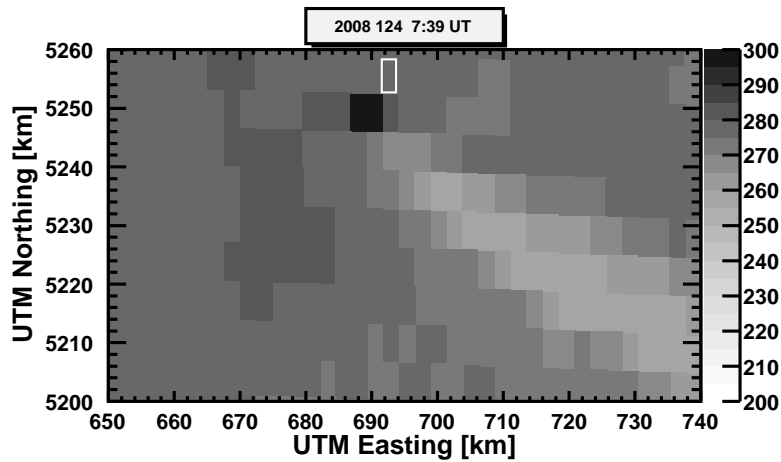}
\par\end{centering}
\caption{ T2 map of the active Chaiten volcano region. The brightness temperature T2 [K] is displayed in gray scale. The red rectangle is the pixel that should encompass the location of the volcano. A displacement of one pixel is apparent in this image.  \label{fig:cloudmaps-2} }
\end{figure}

To ascertain the magnitude of the coordinate fluctuations we monitored the position of a known IR point source, the Chaiten Volcano during its eruption in May 2008. The Chaiten Volcano is conveniently located in UTM Zone 18 at 
692408 Easting and 5255067 Northing, about 860 km Southwest of the Pierre Auger Observatory. The erupting volcano was  identified with the hottest pixel in maps of the brightness temperature T2. An example of one such map is shown in Figure \ref{fig:cloudmaps-2}. The volcano appears as a black pixel at the top of the image. The ash plume appears as a much cooler linear feature extending from the volcano pixel to the Southeast. 
 
\begin{figure}[!h]
\begin{centering}
\includegraphics[width=15.2cm,height=7.0cm,keepaspectratio]{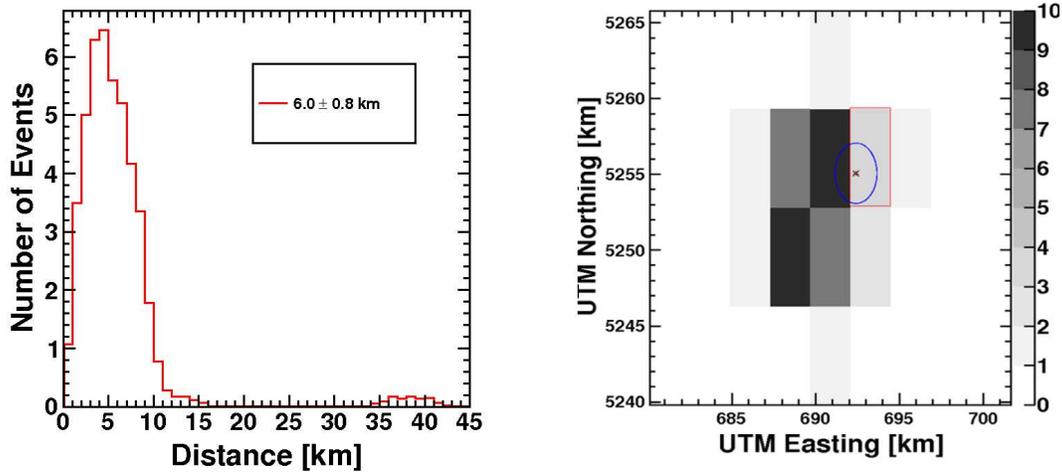}
\par\end{centering}
\caption{Left: A histogram of the distance between the observed hottest spot and the actual location of the centre of the caldera of the volcano. The mean of the histogram is displayed.  For this histogram, each pixel was split into 40 parts with equal area and the caldera was considered as 2.5 km by 4 km. Right: a 2D histogram of the hottest pixel in each of the 44 images is displayed. The mark corresponds to the geographical position of the volcano. The ellipse is a representation of  the caldera of the volcano.\label{fig:cloudmaps-2a} }
\end{figure}

 We were able to identify the hot spot corresponding to the volcano in 45 satellite images from 12 nights. A histogram of the separation between the observed hot spots and the published location of the volcano is displayed in the left panel of Figure \ref{fig:cloudmaps-2a}. The mean displacement of the volcano from its nominal coordinates is about 6 km. Not all of this is attributable to satellite pointing uncertainty as the position of the erupting vent probably changed by several kilometers during the course of the observations. The vent had an equal probability of occurring anywhere within the 2.5 km by 4 km caldera encompassing the volcano. In the right panel of Figure \ref{fig:cloudmaps-2a} we have histogrammed the two-dimensional locations  of the observed hot spots. The hot spot appears to move mainly between two adjacent pixels along a diagonal axis. The caldera happens to be oriented along the same axis. We infer from this that the vent location is fluctuating within the boundaries of the caldera.  
Given this contribution to the uncertainty of the position of the hot spot, we can only use this study to set an upper limit on the spatial accuracy of the satellite pixels. The accuracy is certainly better than the 6 km figure suggested by this study. 

\section{Summary}

We have shown that it is possible to calculate the cloud probabilities 
based on infrared satellite information alone. We expanded this method to assign cloud probabilities to each of the 539 pixels making up the scene based on comparisons between a specific central pixel and ground-based vertical laser shots and under the assumption that the geographical and meteorological conditions of the other 538 pixels are similar to the ones of the central pixel.

As an application of our method, cloud probability maps for the Auger Observatory are generated. These maps are commonly available every 30 minutes during the night.

Our method could be useful for other ground-based and space-based observatories in the region of GOES 12 and GOES 13, specially since the data is publicly available. For observatories in North America the monitoring is even better, since the satellite images are available twice as frequently. 

\section*{Acknowledgments}
The successful installation, commissioning, and operation of the Pierre Auger Observatory
would not have been possible without the strong commitment and effort
from the technical and administrative staff in Malarg\"ue.

We are very grateful to the following agencies and organizations for financial support: 
Comisi\'on Nacional de Energ\'ia At\'omica, 
Fundaci\'on Antorchas,
Gobierno De La Provincia de Mendoza, 
Municipalidad de Malarg\"ue,
NDM Holdings and Valle Las Le\~nas, in gratitude for their continuing
cooperation over land access, Argentina; 
the Australian Research Council;
Conselho Nacional de Desenvolvimento Cient\'ifico e Tecnol\'ogico (CNPq),
Financiadora de Estudos e Projetos (FINEP),
Funda\c{c}\~ao de Amparo \`a Pesquisa do Estado de Rio de Janeiro (FAPERJ),
Funda\c{c}\~ao de Amparo \`a Pesquisa do Estado de S\~ao Paulo (FAPESP),
Minist\'erio de Ci\^{e}ncia e Tecnologia (MCT), Brazil;
AVCR AV0Z10100502 and AV0Z10100522, GAAV KJB100100904, MSMT-CR LA08016,
LG11044, MEB111003, MSM0021620859, LA08015, TACR TA01010517 and GA UK 119810, Czech Republic;
Centre de Calcul IN2P3/CNRS, 
Centre National de la Recherche Scientifique (CNRS),
Conseil R\'egional Ile-de-France,
D\'epartement  Physique Nucl\'eaire et Corpusculaire (PNC-IN2P3/CNRS),
D\'epartement Sciences de l'Univers (SDU-INSU/CNRS), France;
Bundesministerium f\"ur Bildung und Forschung (BMBF),
Deutsche Forschungsgemeinschaft (DFG),
Finanzministerium Baden-W\"urttemberg,
Helmholtz-Gemeinschaft Deutscher Forschungszentren (HGF),
Ministerium f\"ur Wissenschaft und Forschung, Nordrhein-Westfalen,
Ministerium f\"ur Wissenschaft, Forschung und Kunst, Baden-W\"urttemberg, Germany; 
Istituto Nazionale di Fisica Nucleare (INFN),
Ministero dell'Istruzione, dell'Universit\`a e della Ricerca (MIUR), Italy;
Consejo Nacional de Ciencia y Tecnolog\'ia (CONACYT), Mexico;
Ministerie van Onderwijs, Cultuur en Wetenschap,
Nederlandse Organisatie voor Wetenschappelijk Onderzoek (NWO),
Stichting voor Fundamenteel Onderzoek der Materie (FOM), Netherlands;
Ministry of Science and Higher Education,
Grant Nos. N N202 200239 and N N202 207238, Poland;
Portuguese national funds and FEDER funds within COMPETE - Programa Operacional Factores de Competitividade through 
Funda\c{c}\~ao para a Ci\^{e}ncia e a Tecnologia, Portugal;
Romanian Authority for Scientific Research ANCS, 
CNDI-UEFISCDI partnership projects nr.20/2012 and nr.194/2012, 
project nr.1/ASPERA2/2012 ERA-NET and PN-II-RU-PD-2011-3-0145-17, Romania; 
Ministry for Higher Education, Science, and Technology,
Slovenian Research Agency, Slovenia;
Comunidad de Madrid, 
FEDER funds, 
Ministerio de Ciencia e Innovaci\'on and Consolider-Ingenio 2010 (CPAN),
Xunta de Galicia, Spain;
The Leverhulme Foundation,
Science and Technology Facilities Council, United Kingdom;
Department of Energy, Contract Nos. DE-AC02-07CH11359, DE-FR02-04ER41300, DE-FG02-99ER41107,
National Science Foundation, Grant No. 0450696,
The Grainger Foundation USA; 
NAFOSTED, Vietnam;
Marie Curie-IRSES/EPLANET, European Particle Physics Latin American Network, 
European Union 7th Framework Program, Grant No. PIRSES-2009-GA-246806; 
and UNESCO.

We would like to thank the former Michigan Tech students: Nathan Kelley-Hoskins, Kyle Luck and Arin Nelson for their important contribution to the development of this paper. We would like to thank NOAA for the GOES satellite data that we freely downloaded from their website. Also, we would like to mention in these acknowledgments Dr. Steve Ackerman and Dr. Tony Schreiner for very valuable conversations.\\

\bibliography{befcloud_paper}
\bibliographystyle{elsarticle-num}

\clearpage

\end{document}